\documentclass[intlimits,twoside,a4paper]{article}
\usepackage{graphicx,psfrag}
\usepackage[T2A]{fontenc}
\usepackage[cp1251]{inputenc}
\usepackage{soul}
\usepackage[eqsecnum]{cmpj3}

\articletype{Rapid Communication}
\RequirePackage{color}

\issue{2021}{24}{2}{24001}
\doinumber{10.5488/CMP.24.24001}
\title[A mechanism for evolution of the physical concepts network]%
{A mechanism for evolution of the physical concepts network}
\author[V. Palchykov, M. Krasnytska, O. Mryglod, Yu. Holovatch]{V. Palchykov\orcid{0000-0002-5954-653X}\refaddr{label2}, M. Krasnytska\orcid{0000-0002-0464-5741}\refaddr{label1,label2},
O. Mryglod\orcid{0000-0003-4415-7061}\refaddr{label1,label2}, Yu. Holovatch\orcid{0000-0002-1125-2532}\refaddr{label1,label2,label3}}

\addresses{
\addr{label1}Institute for Condensed Matter Physics, National Acad. Sci. of Ukraine, 79011 Lviv, Ukraine 
\addr{label2}$\mathbb{L}^4$ Collaboration \& Doctoral College for the Statistical Physics of Complex Systems, 
Leipzig-Lorraine-Lviv-Coventry, Europe 
\addr{label3}Centre for Fluid and Complex Systems, Coventry University, Coventry, 
CV1 5FB, United Kingdom}


\date{Received May 1, 2021, in final form June 1, 2021}
\begin{document}
\maketitle
\begin{abstract}
We suggest an underlying  mechanism that governs the growth of a network of concepts,
a complex network that reflects the connections between different scientific 
concepts based on their co-occurrences in publications. To this end,
we perform empirical analysis of a network of concepts based on the preprints in physics
submitted to the \texttt{arXiv.org}. We calculate the network characteristics and show that they
cannot follow as a result of several simple commonly used network growth
models. In turn, we suggest that a simultaneous account of two factors, i.e., growth by
blocks and preferential selection, gives an explanation of
empirically observed properties of the concepts network. Moreover, the observed structure
emerges as a synergistic effect of these both factors: each of them alone does
not lead to a satisfactory picture.
\printkeywords
%
%\pacs 
\end{abstract}

Networks of concepts, i.e., semantic networks that reflect the relations between concepts in a
certain domain are ubiquitously met in different spheres of modern life \cite{Sowa1992}. Their importance
is both due to the fundamental reasons and numerous applications, ranging from ontologies in computer
and information science \cite{Fontoura2006} to visual knowledge maps that serve 
as an aid showing where to look for a certain knowledge \cite{vanEck2009}.
Such networks are of particular interest for the logology --- `science of science',
that aims at quantitative understanding of the origins of scientific discovery and creativity, its structure 
and practice \cite{Zeng17,Wang21}.
Scientific papers are an ideal source to investigate such processes, providing
validated and open results of scientific creativity that are recorded in text formats 
and supplied by numerous supporting pieces of information. 
A common approach to the quantitative description of the knowledge structure is via the analysis
of its projections to semantic spaces for different domains, see e.g., \cite{Krenn2020} and references therein. The latter  can be modelled as complex networks based on topic-indicating labels. To give a few examples,
one can mention here the networks of papers in physics that co-used PACS\footnote{Physics and Astronomy Classification Scheme}
numbers \cite{Herrera2010,Pan2012}, biomedical papers that co-mentioned the same chemical entities  \cite{Foster2015},
papers in cognitive neuroscience \cite{Beam2014} and in quantum physics \cite{Krenn2020}
with co-occurence of predefined concepts, Wikipedia pages devoted to mathematical theorems 
\cite{Silva2010}, etc.  In all the above cases, complex network 
formalism enables quantitative analysis of similarities  between different entities 
which are typically considered as indicators of topical relatedness and, 
therefore, as projections of knowledge.

Besides, the networks discussed above rise as an outcome of a dynamical process at which a
new knowledge is acquired. Innovations themselves can be interpreted as an emergence 
of new concepts or new relations between the existing ones
\cite{Iacopini2018,Uzzi2013,Palchykov20}. Modelling such process is a challenging task both for 
its fundamental relevance and numerous practical implementations.
The process of a scientific discovery itself is governed by the structure of 
scientific knowledge. At the same time, it leads to changes in this structure:
in other words, they dynamically update each other. Presence of such a co-evolution
is a typical feature of any complex system \cite{Thurner17,Holovatch17} and is reflected, in particular,
in the growth dynamics of the underlying complex networks of terms, keywords, labels or tags 
that become co-chosen from some predefined semantic space. 
Modelling such complex networks, along with their empirical analysis,
is a challenging task that provides a deeper understanding of their growth mechanisms
\cite{Iacopini2018,Cattuto2009,Rzhetsky2015}.

In this Letter, we suggest an underlying  mechanism that governs the growth of a 
network of concepts originating from the texts of preprints in physics
submitted to e-print repository \texttt{arXiv} \cite{arxiv}. First, we  perform an empirical analysis 
of this network and calculate its topological characteristics. We 
discuss the main network features and show that simultaneous account of two factors i.e., growth by
blocks and preferential selection, gives an explanation of
empirically observed properties. A detailed account of our analysis
is to be published elsewhere \cite{Palchykov21b}.

We used the vocabulary of scientific concepts in the domain of physics that has been collected by the \texttt{ScienceWISE.info} platform \cite{Sciencewise} and refined by continuous updates by expert evaluations. The resulting ontology includes such concepts as \texttt{Ferromagnetism}, \texttt{Quantum Hall Effect}, \texttt{Renormalization group}, and thousands of others. To our knowledge, currently such a vocabulary is the most comprehensive vocabulary 
of this type in the domain of physics. The sample of articles we analysed consists of 36,386 entities submitted to \texttt{arXiv} during a single year 2013 that have been assigned to a single category during submission process and is in one-to-one correspondence with the data set being 
analyzed in \cite{Palchykov20,Palchykov16,Palchykov18}. For each of the articles,
a set of its inherent concepts has been defined using the above mentioned vocabulary of concepts.
In this way, we arrived at the data that are conveniently described as a 
a bipartite network consisting of the nodes of 
two types: articles $A_1,A_2,\dots,A_ {\cal N}$ and concepts $C_1,C_2,\dots,C_N$, each $A$-node
is linked to those $C$-nodes that represent its inherent concepts. While the
properties of the bipartite network and its one-mode projection into the space of articles
were analysed in \cite{Palchykov16,Palchykov18}, here we concentrate
on its one-mode projection into the space of concepts. Now, all $C$-nodes that were connected to the same $A$-node
enter the network as a complete graph or clique.  Hereafter, such a one-mode projection is called a \emph{network of concepts} and is a subject of empirical analysis and modelling.

\begin{table*}[!t]
%\begin{ruledtabular}
\caption{Some features of networks of concepts addressed in our study. An empirically observed network (first line) is compared with 
    three different models discussed in the paper: Erd\H{o}s-R\'enyi,  Barab\'asi-Albert, and growth by blocks with preferential
    selection (GBPS).  The following features are shown: the number of nodes $N$, number of links $L$, density of links $\rho$,
    average node degree $\langle{k}\rangle$, its standard deviation $\sigma$ and maximal value $k_{\rm max}$,
    assortativity mixing by degrees $r$,
    mean clustering coefficient $\langle{c}\rangle$ and global transitivity $T$. 
    }
\vspace{2ex}
    \centering
    \begin{tabular}{|l|c|c|c|c|c|c|c|c|c|}
    \hline
         & $N$ & $L,\times10^6$ & $\rho$ & $\langle{k}\rangle$ & $\sigma$ & $k_{\rm max}$ & $r$ & $\langle{c}\rangle$ & $T$ \\ \hline\hline
        empirical  & 11853 & 5.38 & 7.66\% & 908 & 1146 &  $9970$ & $-0.28$  &  0.74  & 0.38 \\\hline
        Erd\H{o}s-R\'enyi  & 11853 & 5.38 & 7.66\% & 908 &   29 & 1023 &  0.00  &  0.08  & 0.08  \\
        Barab\'asi-Albert        & 11853 & 5.38 & 7.66\% & 908 &  568 & 3875 &  0.01  &  0.15  & 0.15  \\
        \hline
         GBPS    & 11554 &  1.50 &  2.25\% &  260 &   788 & 7603 & $-0.62$  &  0.95  & 0.12 \\
\hline
    \end{tabular}
    \label{tab1}
%\end{ruledtabular}
\end{table*}
The main characteristics of the network of concepts constructed based on the data described in the 
former paragraph are given in the first line (denoted as `empirical') of table~\ref{tab1}. There, out 
of many network indicators, we display those that describe the most typical features addressed below. 
In particular, the empirically observed network of concepts is very dense:
the density of links $\rho = 2L/N(N-1) = 7.66\%$. This number indicates that concepts are densely 
connected within a considered discipline: the authors 
who conduct research in physics, extensively use common terminology.
One of the consequences is the high value of the mean node degree. 
Standard deviation of the node degree distribution indicates a high level of inhomogeneity among 
concept co-occurrence statistics. This can be also observed from the skewed shape of the histogram 
of node degree values $N(k)$ as shown in figure~\ref{fig1}{\bf a} by grey discs.
The tail of the histogram may be visually compared with a power-law function $k^{-\gamma}$ with an exponent close to $\gamma=1$.
While this empirical network cannot be formally classified as the so-called \emph{dense network}  \cite{Allaei06,Bonifazi09,Zhou11,Courtney18}, it is much denser compared to other real networks \cite{Palchykov21b}. 
Similar shapes of node degree distributions were found and declared to be robust 
for a few other analogous empirical networks \cite{Cattuto2009,Rzhetsky2015}.
Negative value of the assortative mixing by degrees $r=-0.28$, defined as Pearson 
correlation coefficient between node degrees on both ends of links in the network,
indicates that in the network of concepts, the high-degree nodes attract low-degree 
ones of a high extent. 
The presence of connectivity patterns is featured by comparatively high values
of the mean clustering coefficient $\langle{c}\rangle$ and global transitivity $T$
(cf. $\langle{c}\rangle=T=1$ for the complete graph and $\langle{c}\rangle=T=0$ for a tree).  For a node $i$ of
degree $k_i>1$, the clustering coefficient is a ratio of existing links $m_i$
between its neighbouring nodes to all possible connections between them, 
$c_i={2m_i}{[k_i(k_i-1)]^{-1}}$. In turn, the global transitivity  $T$ is defined as 
a ratio between the number of closed triplets in the network and the total 
number or network triplets \cite{luce1949}. The difference between the two 
values, $\langle{c}\rangle$ and $T$, indicates specific 
topological features of the network.
With quantitative measures of basic network features at hand,  
let us proceed with modelling a growth process that results in network topology similar to the empirically observed one.

\begin{figure}[t]
\centering{\includegraphics[width=7.7cm]{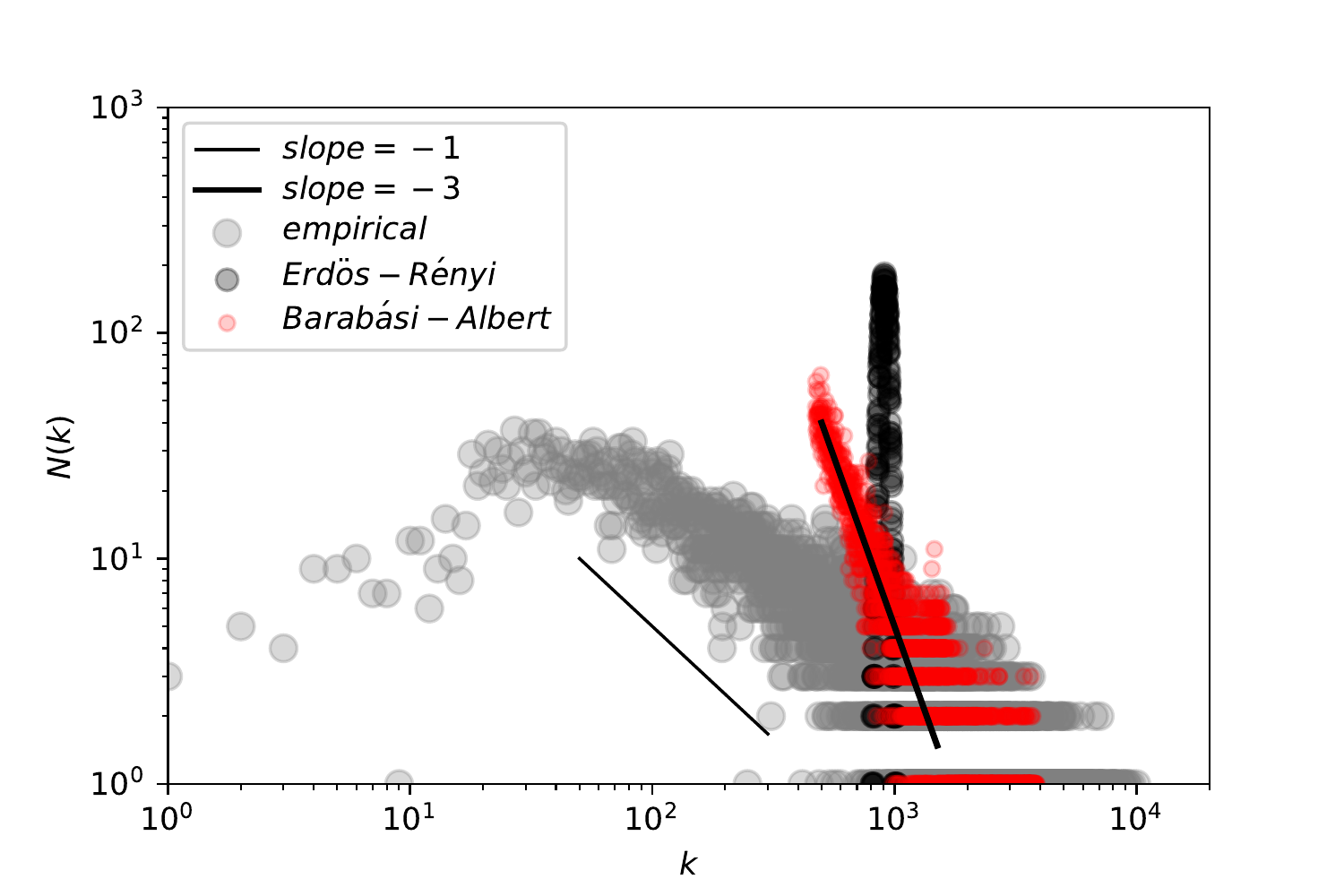}\includegraphics[width=7.7cm]{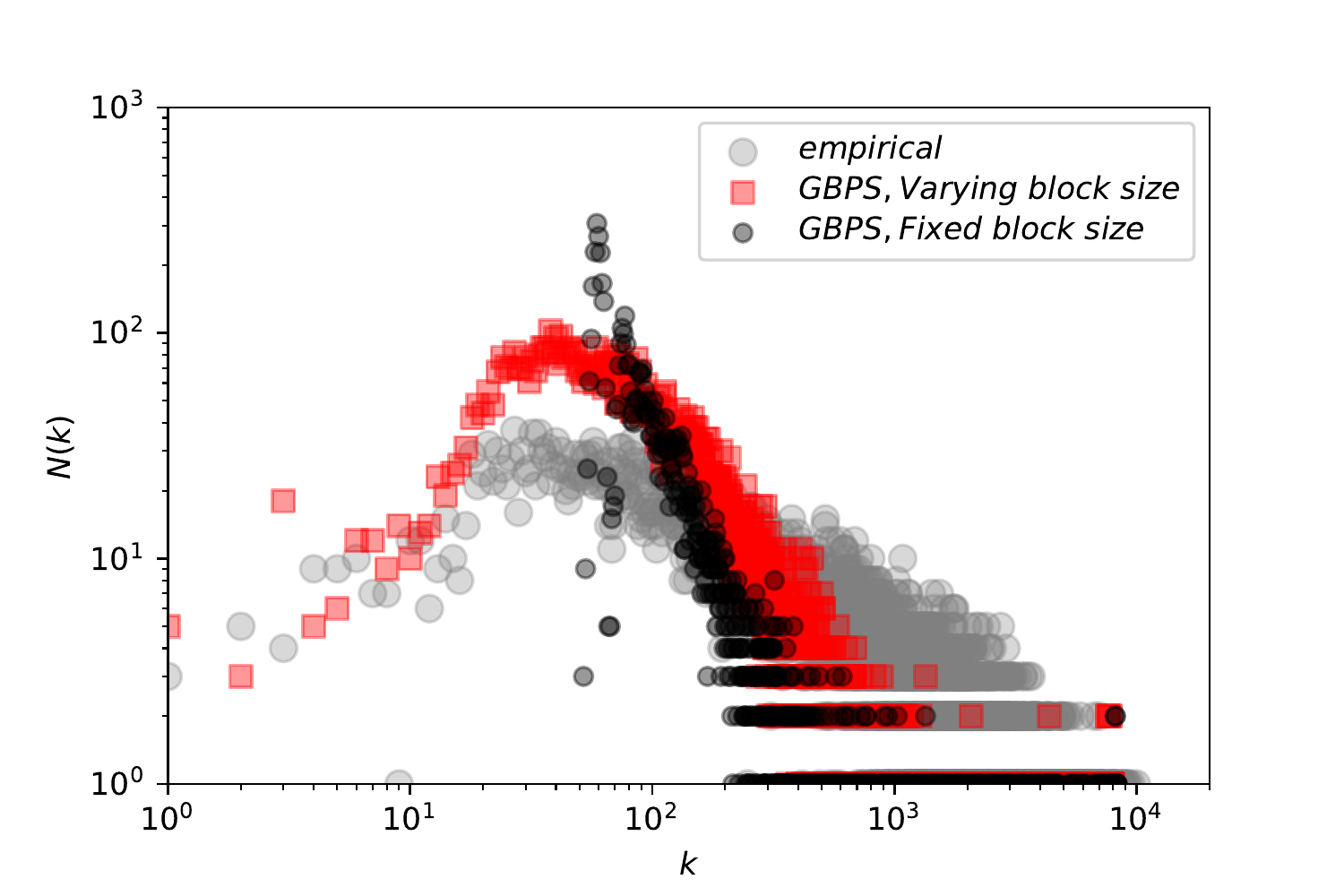}}\\
\centering{{\bf a.} \hspace{20em}  {\bf b.}}
\caption{Node degree histograms $N(k)$ of  networks of concepts addressed in our study. An empirically observed network 
(grey discs) is compared with those generated by Erd\H{o}s-R\'enyi and  Barab\'asi-Albert models (panel {\bf a},
black and red discs, correspondingly) and growth by blocks with preferential
    selection, GBPS (panel {\bf b}, black discs: fixed block size, red squares: varying block size).
\label{fig1}}
\end{figure}

We start with the  Erd\H{o}s-R\'enyi random graph \cite{ER} and  Barab\'asi-Albert preferential attachment \cite{Barabasi1999} 
models. Both models allow us to generate uncorrelated networks with the same number of nodes $N$ and links $L$ as the empirical 
one. Therefore, the density of links $\rho$ and the average node degree $\langle{k}\rangle$ coincide too.
The discrepancies become evident with more in-depth analysis. Results of the network characteristics calculated
for an ensemble average over 100 realizations for each model are shown in the 2nd and 3rd lines of table~\ref{tab1}.
The Erd\H{o}s-R\'enyi random graph is much more
homogeneous than the empirical network: the standard deviation $\sigma$ is almost $40$ 
times smaller than that for the empirical concept network, the 
maximal node degree $k_{\rm max}$ exceeds its average 
value $\langle{k}\rangle$ by $12\%$ only. This may be observed in figure~\ref{fig1}{\bf a}, 
where the corresponding histogram $N(k)$ is shown by black discs.
The Barab\'asi-Albert model, that has growth and preferential attachment as key ingredients,
better reproduces empirical network node degree heterogeneity: 
$k_{\rm max}$ exceeds  $\langle{k}\rangle$ by more than $300\%$,
$\sigma$ almost $20$ times exceeds its value for Erd\H{o}s-R\'enyi graph. 
However, the decay of $N(k)$ is much faster than in the empirical network
(see the red discs in figure~\ref{fig1}{\bf a} and the solid line that corresponds
to $N(k)\sim k^{-\gamma}$ with the Barab\'asi-Albert model decay exponent $\gamma=3$
\cite{Barabasi1999}). The discrepancies are even more pronounced when one considers connectivity 
patterns between nodes of different degrees. Similar to the Erd\H{o}s-R\'enyi graph, the Barab\'asi-Albert 
network is neither assortative nor disassortative, indicating the feature of the empirical network of
concepts that cannot be captured by the models. The other feature 
that is not captured by the models is the difference between the 
average clustering coefficient $\langle{c}\rangle$ and the global transitivity $T$, 
even though the values for the Barab\'asi-Albert model are closer to 
those for the empirical network than the ones for the Erd\H{o}s-R\'enyi network.

To understand the possible mechanisms that lead to the  concept network under consideration, let us develop a model that is capable of reproducing its empirically observed features. Doing so, we do not put as a primary goal 
 reaching a high precision of reproducing the given set of metrics. Rather we are interested in a qualitative
description of the main tendencies in network structure and their explanation by network
generation mechanisms. 
The model of the network evolution that we suggest is based on the 
simultaneous account of two factors: growth by
blocks and preferential selection.
Consider a process with discrete time $t=1\ldots {\cal N}$. 
At each time step, a new article $A_t$ that contains a block of $n_t$ concepts is generated. 
It joins the concept network as a complete graph of $n_t$ nodes.
The article generation consists of two steps: (i) drawing the block size $n_t$ and (ii) 
selecting particular concepts to populate the block. Below, we choose an option when
$n_t$ is drawn from the actual distribution of the number of concepts per article in the 
empirical data set while other options are discussed in \cite{Palchykov21b}. 
  Let us explain step (ii)  more in  detail. When a new article $A_t$ is generated at time $t>1$,
 the already existing data set consists of a set of $t-1$ articles 
 $\mathbb{A}_{t-1}$ and a set of  $N_{t-1}$ different concepts $\mathbb{C}_{t-1}$.
 The new article $A_t$ may contain some of the above $N_{t-1}$ concepts as well as the
{\em novel} concepts that are introduced for the first time. 
Within our model, we fix the probability  of the $i$-th concept 
of article $A_t$ to be a novel one,  $ \pi^{\rm novel}_{t,i}=\nu$. 
Consequently, with probability 
$1-\nu$ a concept of the  generated article is one of the already existing $N_{t-1}$ concepts. 
Moreover, let us consider that the already existing concepts have different chances to be selected 
to populate an article: the more popular the concept is (among the first $t-1$ articles), it is more likely to be 
selected to populate the $t$-th one. We  call such a process a {\em preferential selection}.
The probability $\pi^{\rm exist}_{t,i}(C_j)$ for the concept $C_j$ to be selected 
is proportional to the number of articles ${\cal N}_{t-1}(C_j)$ in which the concept $C_j$ has appeared:
\begin{equation}
    \pi^{\rm exist}_{t,i}(C_j) = \frac{(1-\nu){\cal N}_{t-1}(C_j)}{\sum_{l} {\cal N}_{t-1}(C_l)}, 
    \hspace{1em} C_j\in \mathbb{C}_{t\backslash i-1}\, ,
\end{equation}
where $\mathbb{C}_{t\backslash i-1}$ is the subset of concepts $\mathbb{C}_{t-1}$ 
excluding $i-1$ concepts selected for article $A_t$
and the denominator sums the number of times each concept $C_l$ from the set 
$\mathbb{C}_{t\backslash i-1}$ has appeared in all 
articles.

By the above described  evolution mechanism, the concept network grows by adding cliques to the existing graph. 
At each time $t$, once a new article  $A_t$ of $n_t$ concepts is generated, it enters the concept network as a complete 
graph of $n_t$ nodes and $n_t(n_t-1)/2$  links. Thus, during its evolution, the following processes may be observed 
in a generated concept network: (i) addition of new nodes, (ii) appearance of links between new  nodes and 
between new and already existing nodes, (iii)  appearance of new  links between previously unconnected 
existing nodes, which is important for generation of dense networks. We compare the main features of the network
of concepts generated by the growth by blocks with preferential selection  mechanism in the last line
of table~\ref{tab1}. As for the two previously described  models, we display the values averaged over
an ensemble of 100 network realizations.  The number of articles generated in our simulations was set to 
be exactly the same as the number of articles (${\cal N}=36,386$) in the empirical data set. Fixing the number 
of articles does not ensure that the generated network will have the same number of nodes (concepts). The 
remaining free parameter of the model has been chosen $\nu=8.8\cdot 10^{-3}$ to give a reasonable value
of the number of concepts $N$, see \cite{Palchykov21b} where other concept selection mechanisms were considered.
As one can see from the table, now the modeled network of concepts possesses two features that 
Erd\H{o}s-R\'enyi and Barab\'asi-Albert models failed to reproduce: it is disassortative ($r<0$) and its
mean clustering coefficient and global transitivity differ from each other. The fact that the 
growth by blocks and preferential selection mechanism correctly grasp the main features of
the network of concepts is further supported by the form of the node degree histogram,
as shown by red squares in figure~\ref{fig1}{\bf b}. Now one observes characteristic decays
in the regions of small and large values of $k$. Black discs in the plot show an outcome
of the modified model, when each block of concepts has a fixed size \cite{Palchykov21b} 
that leads to an obvious sharp lower bond.

In the forthcoming publication \cite{Palchykov21b} we will give a more detailed account of the suggested network evolution mechanism along with the analysis of its various modifications. Several conclusions are at hand to finalize this brief report. First of all, one should not go too far in trying to  reach a one-to-one mapping between the features of the empirically observed network of concepts and the modeled one. Indeed, the model which selects new concepts at random, completely neglects their  content-related characteristics. Rather, the goal is to reveal which processes in the network evolution are relevant to  its generic features. As we show in this report, these are the growth by blocks and preferential selection. Moreover, our analysis shows that the observed network structure emerges as a synergetic effect of both of these factors: each of them alone does not lead to a satisfactory picture. The model suggested in this paper may be also of relevance in analysing the generating mechanisms for dense networks which are the subject of ongoing interest \cite{Bonifazi09,Allaei06,Zhou11,Courtney18}. 

This work was supported in part by the National Academy of Sciences of Ukraine, project KPKBK 6541030 (O.M.
\& Yu.H) and by the National Research Foundation of Ukraine, project 2020.01/0338 (M.K.).

%\newpage

%\label{last@page}
%\end{document}
\newpage
\ukrainianpart

\title{╠хїрэ│чь хтюы■Ў│┐ ьхЁхц│ Ї│чшўэшї ъюэЎхяЎ│щ}
\author{┬. ╧ры№ўшъют\refaddr{label2}, ╠. ╩ЁрёэшЎ№ър\refaddr{label1,label2},
	╬. ╠Ёшуыюф\refaddr{label1,label2}, ▐. ├юыютрў\refaddr{label1,label2,label3}}

\addresses{
	\addr{label1}▓эёЄшЄєЄ Ї│чшъш ъюэфхэёютрэшї ёшёЄхь ═└═ ╙ъЁр┐эш, тєы. ╤т║эЎ│Ў№ъюую,  1, 79011  ╦№т│т,  ╙ъЁр┐эр
	\addr{label2}╤яiтяЁрЎ  $\mathbb{L}^4$ i ╩юыхфц фюъЄюЁрэЄiт ``╤ЄрЄшёЄшўэр Ї│чшър ёъырфэшї ёшёЄхь'',
	╦ щяЎiуЦ╦юЄрЁшэуi Ц╦№тiтЦ╩ютхэЄЁi, ктЁюяр
	\addr{label3}╓хэЄЁ яышээшї Єр ёъырфэшї ёшёЄхь, ╙э│тхЁёшЄхЄ ╩ютхэЄЁ│, ╩ютхэЄЁ│, CV1 5FB, ┬хышър ┴ЁшЄрэ│ }
\makeukrtitle

\begin{abstract}
	\tolerance=3000%
	╠ш яЁюяюэє║ью ьхїрэ│чь, ∙ю тшчэрўр║ чЁюёЄрээ  ьхЁхц│ ъюэЎхяЎ│щ --- ёъырфэю┐ ьхЁхц│, ∙ю т│фюсЁрцр║ тчр║ьючт' чъш ь│ц Ё│чэшьш эрєъютшьш ъюэЎхяЎ│ ьш, срчє■ўшё№ эр фрэшї яЁю ┐ї ёя│тяю тє є яєсы│ърЎ│ ї. ╟ Ў│║■ ьхЄю■, ьш тшъюэє║ью хья│Ёшўэшщ рэры│ч ьхЁхц│ ъюэЎхяЎ│щ, юёэютрэ│щ эр яЁхяЁшэЄрї ч Ї│чшъш, чртрэЄрцхэшї эр ёхЁтхЁ \texttt{arXiv.org}. ╠ш ЁючЁрїютє║ью ьхЁхцхт│ їрЁръЄхЁшёЄшъш Єр яюърчє║ью, ∙ю тюэш эх ьюцєЄ№ сєЄш юЄЁшьрэ│ чр фюяюьюую■ ъ│ы№ъюї яЁюёЄшї чруры№эютцштрэшї ьюфхыхщ чЁюёЄрээ  ьхЁхц. ┬ ётю■ ўхЁує, ьш яЁюяюэє║ью юфэюўрёэх тЁрїєтрээ  фтюї ЇръЄюЁ│т: чЁюёЄрээ  сыюърьш Єр яхЁхтрцэшщ тшс│Ё, --- ∙ю фр■Є№ яю ёэхээ  хья│Ёшўэю юЄЁшьрэшї їрЁръЄхЁшёЄшъ ьхЁхц│ ъюэЎхяЎ│щ. ╤яюёЄхЁхцєтрэр ёЄЁєъЄєЁр тшэшър║ тэрёы│фюъ ёшэхЁухЄшўэюую хЇхъЄє юсюї Ўшї ЇръЄюЁ│т --- тЁрїєтрээ  ъюцэюую ч эшї юъЁхью эх фр║ чрфют│ы№эю┐ ърЁЄшэш.

	
\end{abstract}

%\label{last@page}
\end{document}